\documentclass[aoas,preprint]{imsart}

\RequirePackage[OT1]{fontenc}
\RequirePackage{amsthm,amsmath}
\RequirePackage{natbib}
\RequirePackage[colorlinks,citecolor=blue,urlcolor=blue]{hyperref}

\arxiv{arXiv:1903.03223}

\startlocaldefs
\numberwithin{equation}{section}
\theoremstyle{plain}

\endlocaldefs

\usepackage{amsmath}
\usepackage{amsthm}
\usepackage{bbm}
\usepackage{graphicx,psfrag,epsf}
\usepackage{enumerate}
\usepackage{natbib}
\usepackage{mathtools}
\usepackage{url} 
\usepackage{algorithm,float}
\usepackage[noend]{algpseudocode}
\usepackage{caption}
\usepackage{subfigure}
\usepackage{color}
\usepackage{soul}
\usepackage{amsfonts}
\usepackage{indentfirst}

\newcommand{\blind}{1}

\algdef{SnEL}{Inputs}{EndInputs}{1}{\textbf{Inputs:} \State}
\algdef{SnEL}{Outputs}{EndOutputs}{1}{\textbf{Outputs:} \State}
\algdef{SnEl}{Likelihood}{EndLikelihood}{\textbf{Likelihood:}}
\algdef{SnEl}{Initialize}{EndInitialize}{\textbf{Initialize:} \State}
\algdef{SnEl}{Posterior}{EndPosterior}{\textbf{Posterior:}\State}
\algdef{SnEl}{Prior}{EndPrior}{\textbf{Prior:}\State}

\DeclareMathOperator*{\argmax}{arg\,max}
\graphicspath{ {fig/fig_update/} }

\begin{document}

\begin{frontmatter}




\if1\blind
{
 \title{\bf Markov-Modulated Hawkes Processes for Modeling Sporadic and Bursty Event Occurrences in Social Interactions}
\begin{aug}
\author{\fnms{Jing} \snm{Wu}\thanksref{m1}\ead[label=e1]{first@somewhere.com}},
\author{\fnms{Owen G.} \snm{Ward}\thanksref{m1}\ead[label=e3]{second@somewhere.com}},
\author{\fnms{James} \snm{Curley}\thanksref{m3}
\ead[label=e4]{}}
\and
\author{\fnms{Tian} \snm{Zheng}\thanksref{m1,m2}\ead[label=e2]{third@somewhere.com}}

\thankstext{m2}{To whom correspondence should be addressed: \href{tian.zheng@columbia.edu}{tian.zheng@columbia.edu}}

\affiliation{Columbia University\thanksmark{m1} and University of Texas at Austin\thanksmark{m3}}


\end{aug}

 \maketitle
} \fi

\if0\blind
{
 \bigskip
 \bigskip
 \bigskip
 \begin{center}
 {\LARGE\bf Markov-Modulated Hawkes Processes for }\\
 \vspace{0.3cm}
 {\LARGE\bf Sporadic and Bursty Social Interactions}
\end{center}
 
 \medskip
} \fi

\bigskip
\begin{abstract}
 Modeling event dynamics is central to many disciplines. Patterns in observed social interaction events can be commonly modeled using point processes. Such social interaction event data often exhibits self-exciting, heterogeneous and sporadic trends, which is challenging for conventional models. It is reasonable to assume that there exists a hidden state process that drives different event dynamics at different states. In this paper, we propose a Markov Modulated Hawkes Process (MMHP) model for learning such a mixture of social interaction event dynamics and develop corresponding inference algorithms. Numerical experiments using synthetic data demonstrate that MMHP with the proposed estimation algorithms consistently recover the true hidden state process in simulations, while email data from a large university and data from an animal behavior study show that the procedure captures distinct event dynamics that reveal interesting social structures in the real data.
\end{abstract}



\end{frontmatter}


\section{Introduction}

Understanding social interaction dynamics, such as event occurrences with a temporally heterogeneous intensity, has become an important topic in many research disciplines. For example, user behaviors and interactions on social networking platforms and online service providers are of great importance for resource allocation and user experience improvement. Many real-world social interaction event dynamics are sporadic in nature, rife with irregular event-intense intervals, and heterogeneous in event densities. Study shows that group-living animals dynamically shift
their interaction behavior in order to stabilize the social system \citep{williamson2017dynamic}, while the timing of many human actions demonstrates a bursty and heavy-tailed pattern \citep{barabasi2005origin}. It is easy to explain the bursty nature of many event dynamics through the example of an email network: during {\em active} hours, one individual sends an initial email to {\em engage} another individual, which acts as a {\em trigger} that leads to a stream of interactions that follows. These active hours will eventually give way to an {\em inactive} intersession, during which email arrivals have no trigger effects. Such sporadic event dynamics with stochastic {\em inactive-active} transitions can be found in many real-world soical interactions, whose irregularity poses challenges for modeling and understanding of the underlying data generating mechanisms.

\begin{figure}[b!]
 \centering
 \includegraphics[width=\textwidth]{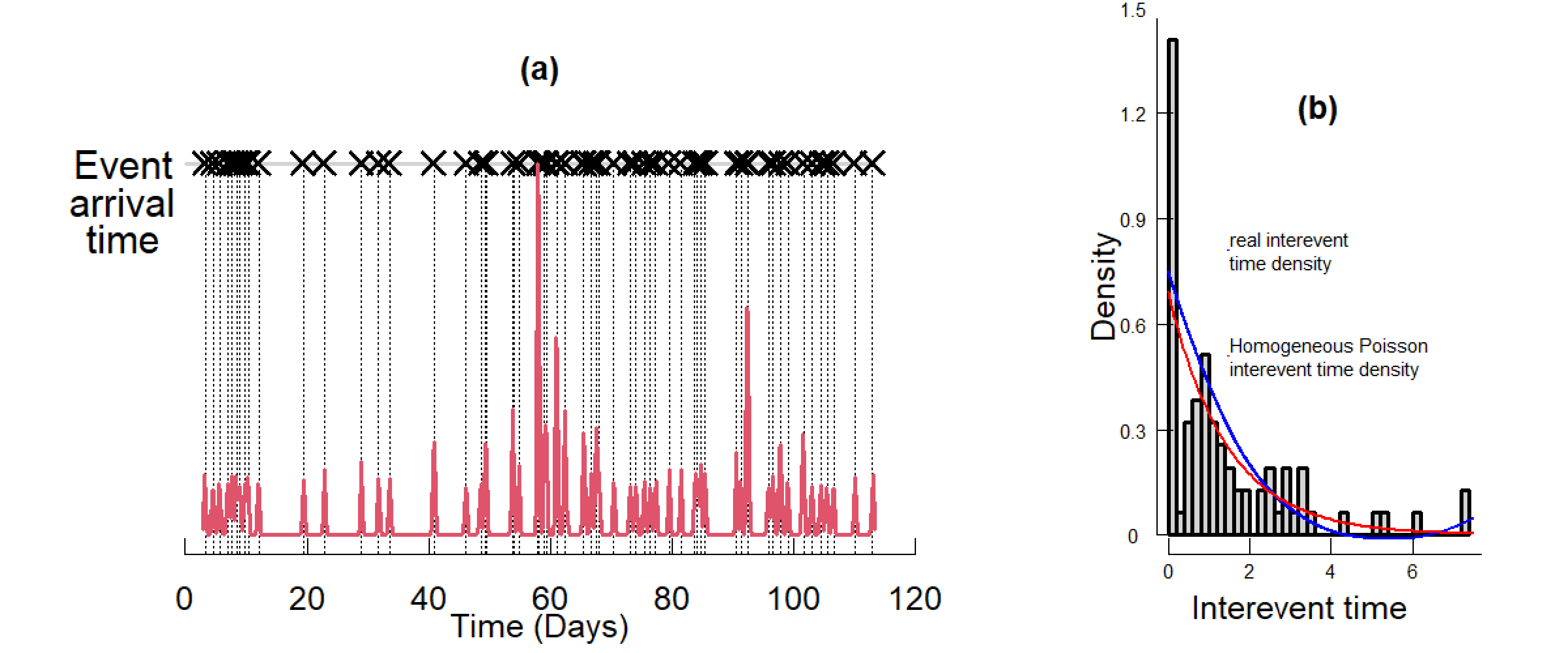}
 \caption{Observed social interactions consisting of emails sent between a pair of users in a large university, as described in \cite{kossinets2006empirical}. \textbf{(a)} describes event arrival times over one semester, with a kernel density estimate for a small bandwidth overlaid, while \textbf{(b)} shows the distribution of interevent waiting times. At certain times events arrive at a very high density. The kernel density (red line) is added to help illustrate these events.}
 \label{fig:ex_email}
\end{figure}


One example of dynamics of this form is the times of emails between users in a large university, as first collected and analysed by \cite{kossinets2006empirical}. Figure~\ref{fig:ex_email} illustrates emails from one user to another in this network over one academic semester (122 days). The data shows distinct periods of activity, such as at the start and end of this time period, along with periods with less frequent interactions. 

Common practices in modeling social interaction event dynamics are generally grouped into two categories. The first type computes aggregated counts of events that are captured in fixed-length time intervals, and then applies time series models for count data (e.g., \citealt{blei2006dynamic}). The second type directly models continuous-time event occurrences via conditional intensity functions (e.g., \citealt{weiss2012multiplicative}). The first approach requires unnecessary aggregation of the data, which inevitably leads to information loss. In most cases, the time series models make assumptions about the true data generating process that are hard to validate. Another common challenge with the first approach is choosing the ``right'' length of time intervals that strikes a balance between count sparsity and information loss, as demonstrated further in Supplement~\ref{supp:data_binning}. This can be especially challenging when arrivals are both sporadic and bursty. The continuous-time approach is a more direct modeling of event dynamics. In recent literature, efforts along this line often treat the observed event arrival times as (heterogeneous) point processes and model their intensity functions with incorporated excitatory triggers \citep{simma2012modeling,zhao2015seismic}, event history \citep{perry2013point}, and/or latent Markov processes \citep{bernardomarkov}.


Poisson processes are the most widely used models for social interaction event arrival times. These models assume a constant event intensity over time and independent event arrivals. For a Poisson process with rate $\lambda$, the waiting time between events follows an exponential distribution with mean $1/\lambda$. However, in practice, the timing of events often follows non-Poisson patterns \citep{barabasi2005origin}. Figure~\ref{fig:ex_email}-(b) displays the distribution of interevent waiting times for the social interactions between a pair of email users shown in Figure~\ref{fig:ex_email}-(a). There are departures from an exponential distribution, with the rate being the maximum likelihood estimate. 
Most notable are departures at the two ends of the distribution that correspond to bursty arrivals and a heavy-tail in waiting times, which have been noted in the literature \citep{barabasi2005origin}.

The Hawkes process \citep{hawkes1971spectra}, a self-exciting process, has been proposed as an alternative to address non-Poisson bursts in event dynamics. In a Hawkes process, at the arrival of an event, the event occurrence intensity is elevated. This boost in event rate is sustained for a short period of time that follows. Hawkes processes have been shown to capture bursty patterns in human activities reasonably well \citep{linderman2014discovering,wang2016coevolutionary}, but are inadequate to address the existence of extended {\em `silent period'} and isolated events. Under a Hawkes process, once an event has occurred, it will always induce an incentive for future events to occur in a short period of time immediately following the ``triggering'' event. However, in reality, long intervals of inactivity or low activity rate between bursts of events are ubiquitous. A more flexible model is needed to address the heavy-tailed distribution of interevent waiting times.

\begin{figure}[ht]
 \centering
 \includegraphics[width=\textwidth]{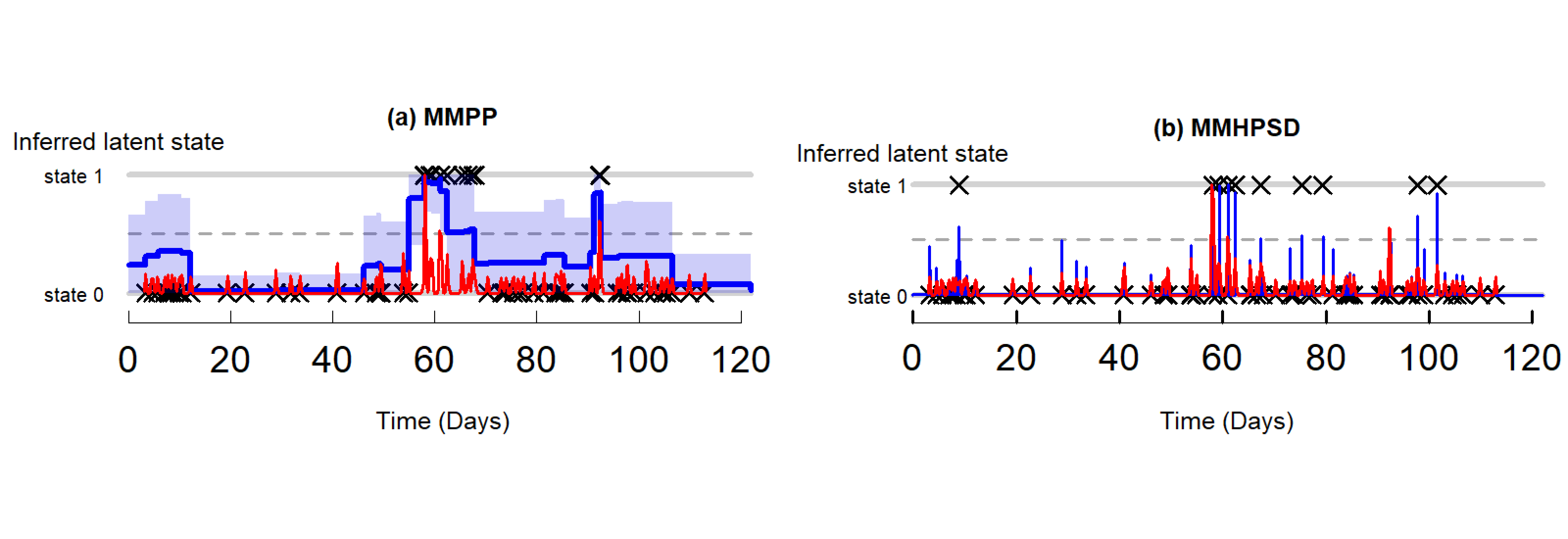}
 \caption{Conventional models applied to the same set of interaction times from a pair of email users as in Figure \ref{fig:ex_email}. {\bf (a)} Interactions with inferred latent states (black crosses) by Markov Modulated Poisson Process (MMPP) \citep{fischer1993markov} with latent trajectory (blue line) and one-standard-error band (blue shade) bounded to be between $0$ and $1$.{\bf (b)} The estimated latent state using Markov Modulated Hawkes Process with stepwise decay (MMHPSD) \citep{wang2012markov}. Overlaid is a KDE with small bandwidth (red line). Both MMPP and MMHPSD failed to detect segments of different types of social interactions, as further demonstrated in
 Supplement~\ref{supp:checking}.}
 \label{fig:ex_email_other_model}
\end{figure}

One model that addresses such heterogeneity in interevent waiting times is the Markov Modulated Poisson Process (MMPP) \citep{fischer1993markov}, which is a doubly stochastic Poisson process with its arrival rate modulated by an underlying Markov process. It assumes that the rate of event arrivals depends on a latent state variable. Conditioning on a given latent state, the arrival of events follows a homogeneous Poisson process. We fit MMPP to data on interactions from the same pair of directed emails in Figure~\ref{fig:ex_email}, see Figure~\ref{fig:ex_email_other_model}-(a). The inferred latent MMPP state (blue line) fails to capture the different event dynamics patterns. In other words, the observed sporadic event dynamics cannot be explained simply by different rates of incidents. Rather, they suggest different levels of {\em temporal dependence}. 


In this paper, we address the modeling of sporadic interevent waiting times, commonly found in real-world social interaction event dynamics. Combining Markov modulation and the Hawkes process, we propose a Markov Modulated Hawkes Process (MMHP) model and develop corresponding inference algorithms. As shown in Figure~\ref{fig:ex_email_other_model}, the interevent time density from real data includes both short ``bursty periods'' and extented ``silent periods'' with isolated events. Our model, on one hand, will address the limitation of the Hawkes process in capturing ``silent periods'' and isolated events. On the other hand, it will extend the flexibility of the MMPP to allow ``bursty periods''.

\citet{wang2012markov} considered a related strategy, where a Hawkes process with step-wise decay was introduced into the MMPP framework (MMHPSD) to model variation in seismicity during earthquake sequences. From a modeling perspective, MMHPSD assumes that each event occurrence creates a constant influence on the intensity function that accumulates with that of other events. This influence is reduced to a lower constant by each event that occurs afterwards. This assumption is not flexible enough to describe certain social event arrivals as it ignores time-decaying effects of previous events as time elapses. The estimation of MMHPSD was implemented using the EM algorithm \citep{dempster1977maximum}, where the M-step computation highly relies on the piece-wise constant assumption. It is therefore difficult to generalize this procedure to the more widely-used exponential kernel for Hawkes processes. The blue line in Figure~\ref{fig:ex_email_other_model}-(b) shows the fitted latent state using the MMHPSD model, where we can see that the latent process is `activated' by nearly every event occurrence, 
classifying several events in the active state. 
When there is no event, the state immediately drops to state 0. 
This is the drawback in using a piece-wise constant intensity function that renders the inference dependent only on local patterns (event versus no events). It fails to detect a stable global latent process that represents a mixture of event dynamics.

Our proposed model considers the Hawkes process with exponential decay, the original more general definition of the Hawkes process. This allows us to efficiently and explicitly model the extent of influence that past events have on the arrival intensity of future events. We derive a novel inference algorithm to solve the computational challenge, providing a close mean-field variational approximation \citep{blei2017variational} of the original likelihood. This novel approximation allows the model inference to be carried out using techniques from the forward-backward algorithm \citep{rabiner1989tutorial} and the Viterbi algorithm \citep{rabiner1989tutorial}, which can be easily generalized to estimate other related models, e.g., MMPP and MMHP with stepwise decay or other kernel functions. This inference procedure can be incorporated into posterior inference under a Bayesian framework that allows us to quantify uncertainty in the model estimates, especially for the latent state process. We evaluate the performance of MMHP using experiments on synthetic data, real email data from \cite{kossinets2006empirical} and mice interaction data from \citet{williamson2016temporal}. MMHP is shown to have excellent model estimation and reliable recovery of the latent states ({\em active} versus {\em inactive}) for synthetic event dynamics with ground truth. MMHP is also better able to capture the dynamics underlying email interactions than previous models. When applied to interaction dynamics among cohorts of male mice, MMHP identifies two types of fighting activity states with different social structures. 



\section{Markov Modulated Hawkes Processes}
\label{model}
We first introduce the proposed MMHP model for sporadic event dynamics given observed event arrival times. We start with the notation that is necessary for our discussion. Then, we lay down the background on point processes in general with a focus on the Hawkes process. Finally, we introduce the proposed MMHP, a latent variable model with Hawkes process modulated by a Markov process. 

\paragraph{Notation for event arrival time data} We consider event arrival time data that consists of all event history up to a {\em final-observation} time $T$: $\mathcal{H}(T) = \{t_m\}_{m=0}^M$, where $t_0=0$, $t_M=T$, and $M$ is the total number of events. The sequence of interevent waiting times is denoted by $\{\Delta t_m := t_m-t_{m-1}\}_{m=1}^{M}$, which is equivalent to the event time $\mathcal{H}(T)$.

\paragraph{Background on point process models} An equivalent representation of a point process $\mathcal{H}(T) = \{t_m\}_{m=0}^M$ is via a {\em counting process}. Let $N(t)$ be a right-continuous point process that records the number of events observed during the interval $[0,t]$. The conditional intensity function given the history up to time $t$, $\mathcal{H}(t)$, is $$\lambda(t|\mathcal{H}(t))=\lim_{\Delta t\to 0} \frac{Pr(N(t+\Delta t)-N(t^-)=1|\mathcal{H}(t))}{\Delta t}.$$ The likelihood function for a sequence of events up to time $T$, $\mathcal{H}(T) = \{t_1<...<t_M\}$, is then 

\begin{equation}\label{eq:likelihood}
\prod_{m=1}^M \lambda(t_m|\mathcal{H}(t_m))\exp\Big\{-\int_{0}^{T} \lambda(u|\mathcal{H}(u)) du\Big\}.
\end{equation}

The {\em Hawkes process} \citep{hawkes1971spectra} is a self-exciting process that can explain bursty patterns in event dynamics. For a univariate model with the widely used exponential kernel \footnote{Other kernel functions such as a powerlaw kernel have also been used in applications in seismology and finance.}, the intensity function is defined as
\begin{equation}
 \lambda(t) = \lambda_1 + \int_{0}^t\alpha e^{-\beta(t-s)} dN_s = \lambda_1 + \sum_{t_m<t}\alpha e^{-\beta(t-t_m)},
\end{equation}
where $\lambda_1 > 0$ specifies the baseline intensity, $\alpha>0$ calibrates the instantaneous boost to the event intensity at each arrival of an event, and $\beta>0$ controls the decay of past events' influence over time. Hawkes processes have successfully been used to model dynamics in fields such as finance \citep{hawkes2018hawkes} and
Neuroscience \citep{linderman2014discovering}.



\paragraph{Modulation by a latent Markov process} In \citet{fischer1993markov}, the original Poisson process was extended to be {\em modulated} by a latent continuous-time Markov chain (CTMC), primarily to address the commonly non-Poisson pattern where the event dynamics alternate between long waiting times and intervals of more intensive events. More specifically, this Markov Modulated Poisson Process (MMPP) model is a doubly stochastic Poisson process whose arrival rate is given by $\lambda_{Z(t)}$. $Z(t)$ is an irreducible Markov process with $R$-states that is independent of the arrival process. When the Markov process $Z(t)$ is in state $r\ (r\in\{1,...,R\})$, arrivals occur according to a Poisson process of rate $\lambda_r$. In this paper, we will consider a
two-state $Z(t)$ that takes values 0 or 1. All the following discussion can be generalized to an $R$-state Z(t).

MMPP assumes a constant event intensity conditioning on the latent state $Z(t)$. For sporadic event dynamics with bursts, isolated incidents and long waiting times, we propose to use Hawkes process to model a self-exciting $\lambda_1(t)$ instead of using a constant rate $\lambda_1$, when the underlying Markov process is in the active state ($Z(t) = 1$). This Hawkes process $\lambda_1(t)$ considers the whole event history up to current time $t$, i.e., $\mathcal{H}(t)$. When $Z(t)=0$, the point process follows homogeneous Poisson process with rate $\lambda_0$. Modulating the intensity function using a latent two-state Markov process $Z(t)$ allows us to extract segments with heterogeneous event dynamics. 
$Z(t)$ is described by an initial probability vector $\mathbb{\delta}:=(1-\delta_0,\delta_0)$, and an infinitesimal generator matrix $Q$, where
\begin{align*}
Q&=
\begin{bmatrix}
q_{1,1} & q_{1,0} \\
q_{0,1} & q_{0,0}
\end{bmatrix}
=
\begin{bmatrix}
-q_{1} & q_{1} \\
q_{0} & -q_{0}
\end{bmatrix}.
\end{align*}

For each $t\geq 0$, there is a probability transition matrix, denoted as $\mathbf{P}(t):=$ $[\mathbf{P}_{ij}(t)]_{i,j\in\{0,1\}}$. Each entry $\mathbf{P}_{ij}(t)$ is defined as the probability that the chain will be in state $j$ at time $u+t$ ($t>0$) given the chain is in state $i$ at time $u$, i.e. for each $u\geq 0$, $$\mathbf{P}_{ij}(t)=P(Z(u+t)=j|Z(u)=i).$$

\begin{figure}[h!]
 \centering
 \includegraphics[width=\textwidth]{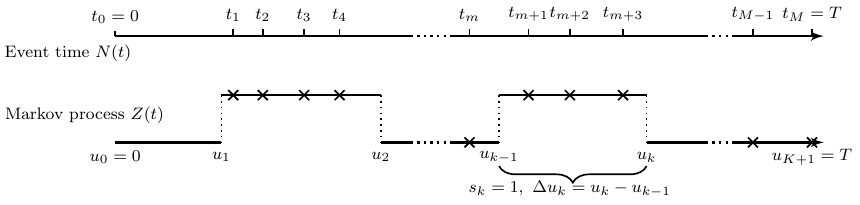}
 \caption{An illustration of the MMHP model.}
 \label{fig:plot}
\end{figure}

The likelihood of the full trajectory of the hidden state $\{Z(t), t\leq T\}:=\mathcal{Z}(T)$ can be written in terms of the following sufficient statistics: initial state $Z(0)$, the number of jumps $K$ and the successive transition time points $\{u_1,...,u_K\}$, given parameters $\delta$, $q_1$ and $q_0$. Further, denote $s_k$ as the state of $Z(t)$ during $[u_{k-1},u_k)$, and $\Delta u_k=u_k-u_{k-1}$, as shown in Figure \ref{fig:plot}. Then the likelihood function is written as
\begin{equation*}
P(\mathcal{Z}(T)|\delta,Q) = \, \delta_0^{I\{Z(0)=0\}}(1-\delta_0)^{I\{Z(0)=1\}} [\prod_{k=1}^K q_{s_k}e^{-q_{s_k}\Delta u_k}\frac{q_{s_k,s_{k+1}}}{q_{s_k}}] e^{-q_{s_{K+1}}\Delta u_{K+1}} . 
\end{equation*}

Let $\Theta$ denote the entire set of parameters, i.e., $\{\lambda_0,\lambda_1, \alpha,\beta,\delta,q_1,q_0\}$.
The complete-data likelihood for $\Theta$, under MMHP, is then
\begin{align}
P(\mathcal{H}(T), \mathcal{Z}(T)|\Theta) = \,P(\mathcal{Z}(T)|\Theta) \times P(\mathcal{H}(T) | \Theta, \mathcal{Z}(T)), 
 \label{eq:lik}
\end{align}
where $P(\mathcal{Z}(T)|\Theta)$ is for the latent Markov process, which is not dependent on the observation $\mathcal{H}(T)$. $P(\mathcal{H}(T)|\Theta,\mathcal{Z}(T))$ is for observed event data conditioning on the latent process. More specifically,
\begin{align}
P(\mathcal{H}(T) | \Theta, \mathcal{Z}(T)) = \prod_{m=1}^M \lambda_{Z(t_m)}(t_m|\mathcal{H}(t_m))\exp{\{-\int_0^T \lambda_{Z(u)}(u|\mathcal{H}(u)) du\}}.
\end{align}


\section{Bayesian inference of MMHP}

\label{inference}

\subsection{Inference of model parameters}
In this paper, we adopt a Bayesian framework for the inference of MMHP. We are interested in the posterior distribution of parameter set $\Theta$, given a proposed prior distribution $\pi(\Theta)$ and observed $\mathcal{H}(T)$. It can be written that, 
\begin{equation}
\label{eq:marginalization}
P(\Theta|\mathcal{H}(T))\propto \pi(\Theta)P(\mathcal{H}(T)|\Theta) = \pi(\Theta)\sum_{\mathcal{Z}(T)}P(\mathcal{H}(T),\mathcal{Z}(T)|\Theta).
\end{equation}
{\color{black} For better computational stability and convergence, in (\ref{eq:marginalization}), we integrate out the full latent state trajectory.}
However, the exact marginalization is computationally infeasible over the entire set of possible full trajectories of continuous-time Markov chains (CTMC), 
$\mathcal{S}:=\{\mathcal{Z}(T)\}$. In practice, we are more interested in the latent 
state at the time of each event (and also the initial state), i.e. 
$\tilde{\mathcal{Z}}(T):=\{Z_0, Z_1, Z_2, ..., Z_M\}$. In order to find an efficient 
approximation of (\ref{eq:marginalization}), we divide the computation into subsets of 
$\mathcal{S}$ with $$\mathcal{S}=\bigcup\limits_{\tilde{\mathcal{Z}}(T)} 
\mathcal{S}_{\tilde{\mathcal{Z}}(T)},\ 
\mathcal{S}_{\tilde{\mathcal{Z}}(T)}:=\{\mathcal{Z}(T)|Z(t_m)=Z_m, 
Z_m\in\tilde{\mathcal{Z}}(T), m=1,...,M\}.$$

{\color{black} Here $\mathcal{S}_{\tilde{\mathcal{Z}}(T)}$ is the set of CTMC trajectories 
that result in
${\tilde{\mathcal{Z}}(T)}$. Given each possible combination of observed event states,
${\tilde{\mathcal{Z}}(T)}$, the entire set $\mathcal{S}$ can 
be decomposed into the union of
$2^{M+1}$ such $\mathcal{S}_{\tilde{\mathcal{Z}}(T)}$ realizations.}

The summation in (\ref{eq:marginalization}) over a CTMC family of infinite dimension is 
then divided into $2^{M+1}$ parts,
\begin{equation}
\label{eq:sum}
\sum_{\mathcal{Z}(T)}P(\mathcal{H}(T),\mathcal{Z}(T)|\Theta) = \sum_{\tilde{\mathcal{Z}}(T)} \sum_{\mathcal{Z}(T)\in\mathcal{S}_{\tilde{\mathcal{Z}}(T)}}P(\mathcal{H}(T),\mathcal{Z}(T)|\Theta).
\end{equation}

For each term in the outer summation of (\ref{eq:sum}), a variational 
approximation solution is proposed in section \ref{inference-variation} 
\citep{cohn2009mean}. Based on these approximated likelihood summands,
we construct an inference procedure for the whole marginalization in 
(\ref{eq:marginalization}), utilizing components of the forward-backward 
algorithm \citep{rabiner1989tutorial}. This allows us to implement an MCMC 
sampling algorithm and derive the posterior distribution of $\Theta$.
The latent Markov process trajectory is then estimated by the Viterbi 
algorithm \citep{rabiner1989tutorial} and interpolated by maximizing the 
likelihood of $\mathcal{Z}(T)$, as outlined in section 
\ref{inference-latent}. {\color{black} Empirical results (not shown)
indicate that this inference procedure, obtaining both
the posterior distribution $\Theta$ and the latent process
trajectory, scales linearly in the number of events observed.}

\subsection{Likelihood approximation}
\label{inference-variation}
Working with each summand in (\ref{eq:sum}), 
\begingroup
\allowdisplaybreaks
\begin{align}
\label{eq:approx}
\sum_{\mathcal{Z}(T)\in\mathcal{S}_{\tilde{\mathcal{Z}}(T)}}P(\mathcal{H}(T),\mathcal{Z}(T)|\Theta) 
&= \sum_{\mathcal{Z}(T)\in\mathcal{S}_{\tilde{\mathcal{Z}}(T)}} P(\mathcal{Z}(T)|\Theta) \times P(\mathcal{H}(T) | \Theta, \mathcal{Z}(T))\notag\\
&= P(\tilde{\mathcal{Z}}(T)|\Theta)\times E_{\mathcal{Z}|\tilde{\mathcal{Z}}} [P(\mathcal{H}(T) | \Theta, \mathcal{Z}(T))].
\end{align}
\endgroup

\paragraph{The calculation of $P(\tilde{\mathcal{Z}}(T)|\Theta))$}
This is the likelihood of the Markov process embedding at event times, which can be calculated as
\begin{align}
\label{eq:approx_1}
P(\tilde{\mathcal{Z}}(T)|\Theta) = \delta_0^{I\{Z_0=0\}}(1-\delta_0)^{I\{Z_0=1\}}\prod_{m=1}^M \mathbf{P}_{Z_{m-1},Z_m}(\Delta t_m).
\end{align}

Following Kolmogorov's forward equation $$\frac{d\mathbf{P}_{ij}(t)}{dt}=\sum_{k\neq j}q_{kj}\mathbf{P}_{ik}(t)-
\left(\sum_{k\neq j}q_{jk}\right)
\mathbf{P}_{ij}(t),$$ {\color{black} we can readily calculate the matrix $\mathbf{P}(t) = \exp{(Qt)}$. When there are two states, it has an explicit form as follows}, 

\begin{align*}
\mathbf{P}(t)&
= \begin{bmatrix}
\mathbf{P}_{11}(t) & \mathbf{P}_{10}(t) \\
\mathbf{P}_{01}(t) & \mathbf{P}_{00}(t)
\end{bmatrix}
=\frac{1}{q_0+q_1}
\begin{bmatrix}
q_0+q_1e^{-(q_0+q_1)t} & q_1-q_1e^{-(q_0+q_1)t} \\
q_0-q_0e^{-(q_0+q_1)t} & q_1+q_0e^{-(q_0+q_1)t}
\end{bmatrix}.
\end{align*}

\paragraph{Variational approximation of $E_{\mathcal{Z}|\tilde{\mathcal{Z}}} [P(\mathcal{H}(T) | \Theta, \mathcal{Z}(T))]$} We need to approximate the marginalization of all possible trajectories of $\mathcal{Z}(T)$ given a $\tilde{\mathcal{Z}}(T)$. More specifically, we approximate the expectation in the following equation,
\begin{align}
\label{eq:approx_2}
E_{\mathcal{Z}|\tilde{\mathcal{Z}}} \left [
P(\mathcal{H}(T) | \Theta, \mathcal{Z}(T))\right] = \prod_{m=1}^M \lambda_{Z_m}(t_m|\mathcal{H}(t_m)) E_{\mathcal{Z}|\tilde{\mathcal{Z}}}
\left[\exp{\{-\int_0^T \lambda_{Z(u)}(u|\mathcal{H}(u)) du\}}\right],
\end{align}
where, for $u\in[0,T]$, 
$$
\lambda_{Z(u)}(u|\mathcal{H}(u)) =\lambda_1(u|\mathcal{H}(u))Z(u)+\lambda_0(1-Z(u)) 
\coloneqq
\lambda(Z(u),u).$$ 
We then derive a variational approximation for
\begin{equation}
\label{eq:integration}
E_{\mathcal{Z}|\tilde{\mathcal{Z}}}
\left[\exp{\left( -\int_0^T \lambda(Z(u),u) du\right)} \right].
\end{equation}

Variational approximation is a general numerical tool for approximating any integral and has been widely used for approximating full posteriors \citep{blei2017variational}. {\color{black} \citet{cohn2009mean} pointed out that inference problems related to CTMC are computationally intractable and it is necessary to construct an approximation. Given the observations of the event times, our problem setting is similar to \citet{cohn2009mean}, except that our observations are point process. Under our model assumptions, we assume the \emph{evidence} of the states is $\tilde{Z}$, e.g., the latent states at the event times. Thus, for our approximation task, we consider the Markov process density family defined in \citet{cohn2009mean}: $$\mathcal{M}_{\tilde{Z}} := \{\mu_z(t), \gamma_{z_1,z_2}(t): 0\leq t\leq T\}.$$ 
Here
\begin{align*}
\mu_z(t)&=P(Z(t)=z),\ z \in \{0,1 \} \\
\gamma_{z_1,z_2}(t)&=\lim_{h\rightarrow 0}\frac{P(Z(t)=z_1, Z(t+h)=z_2)}{h},\
z_1,z_2 \in \{0,1\},
z_1\neq z_2,
\end{align*}
which satisfies $\mu_{z_m}(t_m)=1$, $\mu_z(t_m)=0,z\neq z_m$, and other positive, normalizing and marginal conditions as stated in Definition 1 of \citet{cohn2009mean}. This definition is critical for the validity of the variational distribution as proven by \citet{cohn2009mean}. Although our computation only relies on the definition of $\mu_z(t)$, we include the second moment definition for the completeness of the discussion.

Given $f_{\tilde{Z}} \in \mathcal{M}_{\tilde{Z}}$, consider the integration $E_{f_{\tilde{Z}}}[\exp{(-\int_0^T \lambda(Z(u),u) du})]$. It can be readily approximated by applying Jensen's inequality as follows.

\begingroup
\allowdisplaybreaks
\begin{align}
\label{eq:variational}
  &E_{f_{\tilde{Z}}}[\exp{(-\int_0^T \lambda(Z(u),u) du})] \\
&\geq \exp{(-E_{f_{\tilde{Z}}}[\int_0^T \lambda(Z(u),u) du}]) 
\notag\\
& = \exp{(-\int_\Sigma f_{\tilde{Z}}(\sigma)\int_0^T 
\lambda(Z(u),u) du d\sigma \notag)}\\
& = \exp{\left(-\int_0^T \sum_z\int_\Sigma 
f_{\tilde{Z}}(\sigma)I_{Z(u)=z}\lambda(z,u) d\sigma du\notag 
\right)}\\
& = \exp{\left(-\int_0^T \sum_z\lambda(z,u) \mu_z(u) du\notag 
\right)}.
\end{align}
\endgroup

Following Theorem 6 of \citet{cohn2009mean}, 
(\ref{eq:integration}) can 
be approximated by the integral (\ref{eq:variational}) evaluated at the
$f_{\tilde{Z}}$ that minimizes the Kullback-Leibler divergence 
\citep{kullback1951information} from the process of interest given 
$\tilde{\mathcal{Z}}(T)$. 
$f_{\tilde{Z}}$ satisfies the condition that,
$$
\mu_z(u)=\frac{\mathbf{P}_{z_{m-1},z}(u-t_{m-1})\mathbf{P}_{z,z_{m}}(t_m-u)}{\mathbf{P}_{z_{m-1},z_{m}}(\Delta t_m)}, \mbox{ for }u\in[t_{m-1},t_m).$$
}


{
\color{black} 
Given the formulation of $\mu_z(u)$, we are able to calculate the integration in the (\ref{eq:variational})
\begin{align}
\label{eq:mu}
&\int_{t_{m-1}}^{t_m} \sum_z\lambda(z,u) \mu_z(u) du \\
&=\sum_z \int_{t_{m-1}}^{t_m} \lambda(z,u) \frac{\exp{(Q(u-t_{m-1}))}_{z_{m-1},z}\exp{(Q(t_m-u))}_{z,z_{m}}}{\exp{(Q\Delta t_m)}_{z_{m-1},z_m}} du.\notag
\end{align}

Thus, we can calculate the summand in (\ref{eq:sum}) by combining (\ref{eq:approx}), (\ref{eq:approx_1}), (\ref{eq:approx_2}) (\ref{eq:variational}) and (\ref{eq:mu}). 
}
This variational approximation allows marginalizing the likelihood over $\left\{  {\mathcal{Z}(T)\in\mathcal{S}_{\tilde{\mathcal{Z}}(T)}}
\right\}$ to be computationally tractable. Given this intermediate approximated result, which will be denoted as
\[
\tilde{P}_\Theta(\mathcal{H}(T),\tilde{\mathcal{Z}}(T)):=\sum_{\mathcal{Z}(T)\in\mathcal{S}_{\tilde{\mathcal{Z}}(T)}}P(\mathcal{H}(T),\mathcal{Z}(T)|\Theta),\]
we will then marginalize over $\tilde{Z}(t)$. This can be carried out in linear time by using the forward component of the forward-backward algorithm \citep{rabiner1989tutorial}, which is described in Appendix~\ref{appendix:for_back}.

\paragraph{Bayesian inference of MMHP using MCMC}
We impose weakly informative priors for the model parameters: $\delta_0\sim U(0,1)$, $\alpha \sim \mbox{N}(0,5)$, $\beta\sim\log\mbox{N}(0,0.5)$, $\lambda_1\sim\log\mbox{N}(0,1)$, $\lambda_0<\lambda_1$, $q_0, q_1 \sim\log\mbox{N}(-1,1)$. $\lambda_1$ is baseline intensity for the Hawkes process, which is greater than the rate in state 0, $\lambda_0$, in order to address model identifiability issue. As part of the MCMC sampler, we incorporate the above likelihood approximation algorithm and forward algorithm as in Appendix~\ref{appendix:for_back} to obtain posterior draws of the parameters. See Algorithm \ref{algo:1} in
Appendix~\ref{appendix:post} for more details. Computation was carried out in Stan \citep{guo2014rstan}.

\subsection{Inference of the latent process}
\label{inference-latent}
Given a posterior draw $\hat{\Theta}$, we may infer the most likely sequence of hidden states, $\hat{z}_{1:M}$, corresponding to the observed events, $\hat{z}_{1:M}$. We apply the Viterbi algorithm \citep{forney1973viterbi}, which maximizes the conditional probability: $P(Z_{1:M}=\hat{z}_{1:M}|\Theta, \Delta t_{1:M})$, as shown in Algorithm \ref{algo:2} in Appendix~\ref{appendix:vit}. Then, the full latent trajectory $\mathcal{Z}(T)$ given $\hat{\Theta}$ and $\hat{z}_{1:M}$ is interpolated by maximizing the likelihood of there being no event between two observed events, given the estimated states at these two events' times and the parameter estimate, i.e., $P(Z(t)|Z_{m}=\hat{z}_{m},Z_{m+1}=\hat{z}_{m+1},\hat{\Theta}, \Delta t_{1:m+1}), t\in[t_m,t_{m+1})$. Based on the posterior latent trajectories corresponding to a sample of the posterior draws of $\hat{\Theta}$, $\mathcal{Z}(T)$ can be estimated by their majority vote at each $t$.

\section{Experiments}
\label{experiments}

\subsection{Experiments on synthetic data} 

To evaluate the validity of our proposed algorithms for estimating MMHP, we simulate event arrival times from a generative MMHP model and explore parameter recovery. The simulation of point processes is based on the {\em thinning algorithm} \citep{ogata1981lewis}, a common approach for simulating inhomogeneous Poisson processes.

\paragraph{Estimation of model parameters} Given one fixed set of parameters $\Theta$, we simulate
{\color{black} $S=100$} independent sets of synthetic MMHP processes, each with an independent latent Markov process, and let them run to $M=50$ events.

Model estimation was carried out using Algorithm~\ref{algo:1} with MCMC sampling. We run four parallel chains with random initial values and 1000 iterations per chain, using the first half of each chain for burn-in. 
For each parameter we obtain $\hat{R}_{\mbox{max}}<1.1$ \citep{gelman2013bayesian}. 
Since the posterior distributions are highly skewed, we calculated simulation-efficient Shortest Probability Intervals (SPIn) \citep{liu2015simulation} for the parameter estimates. For our $95\%$ posterior probability interval using SPIn, the coverage rates of true values are all above $95\%$.

\begin{figure}[h!]
 \centering
\includegraphics[width=\linewidth]{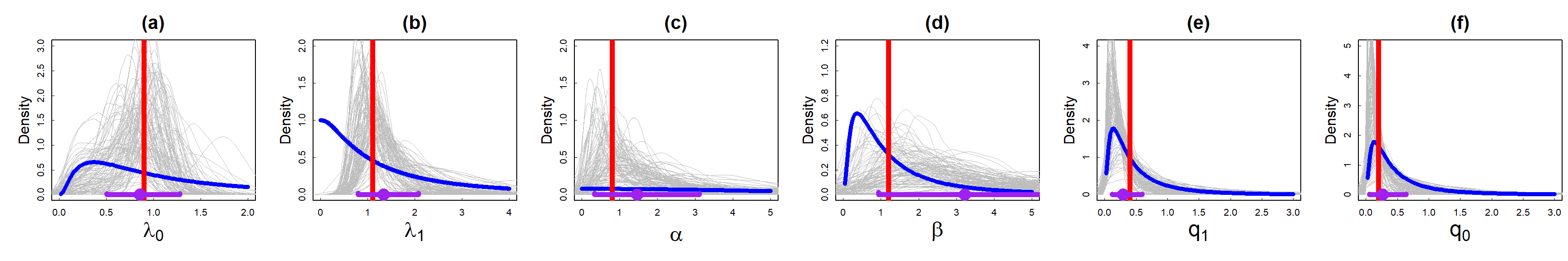}
 \caption{Estimation of the parameters for simulated MMHPs. Red lines: ground truth; grey lines: posterior distributions; purple segments: 95\% Shortest Probability Intervals (SPIn) \citep{liu2015simulation} based on the sampling distribution; blue lines: prior distributions.}
 \label{fig:sim_par}
\end{figure}

We further examined the sampling distribution of the posterior means from the $S=100$ independent simulated data sets. Figure~\ref{fig:sim_par} displays 100 estimated posterior distributions for each model parameters in grey lines. The purple points and horizontal bars are the average of these posterior mean estimates and their shortest 95\% probability interval. The blue lines represent prior distributions and red vertical lines indicate the ground truth. It shows that the sampling distribution of our estimates is centered at the true value with reasonable precision. 

\paragraph{Estimation of the latent Markov process} Using synthetic data, we evaluate how well our proposed algorithm recovers the latent state process. We first simulate one fixed instance of the latent Markov process $\mathcal{Z}(T)$, and then generate {\color{black}$S=100$}
sets of event arrival times given this fixed $\mathcal{Z}(T)$. Numerical experiments are conducted under different lengths of the latent Markov process, where the average numbers of events are $M = 50$, $100$, $200$ and $500$. 

\begin{figure}[ht]
 \centering
 \includegraphics[width=0.98\textwidth]{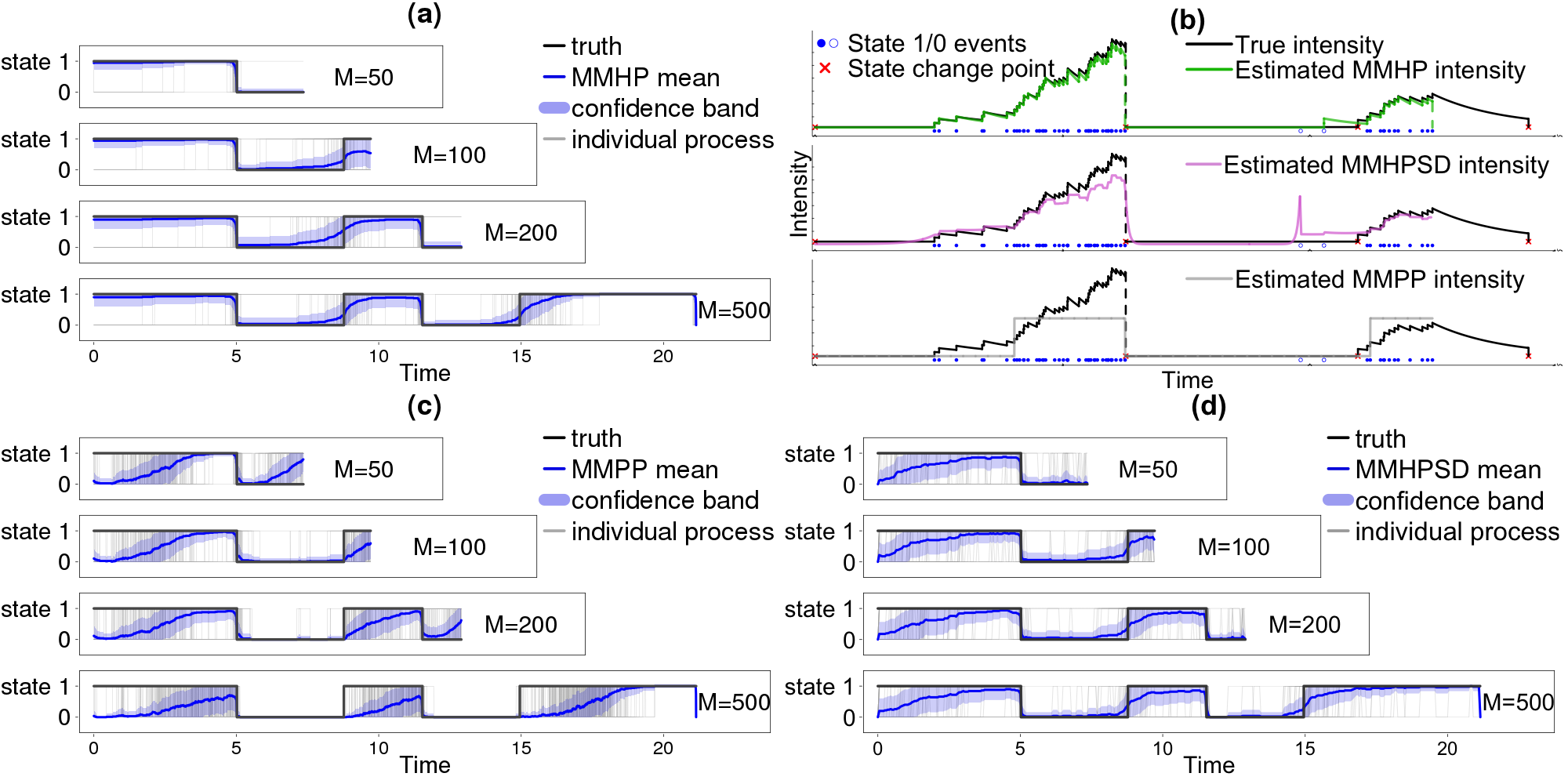}
 \caption{Estimation of the latent Markov process {\bf (a)} Estimated latent trajectory using the proposed algorithm. {\bf (b)} Event intensity for one synthetic process. {\bf (c)} Estimated latent trajectory using MMPP. {\bf (d)} Estimated latent trajectory using MMHPSD.}
 \label{fig:sim_state}
\end{figure}

Following the inference procedure described in section \ref{inference-latent}, we show the estimates of the latent Markov process in Figure \ref{fig:sim_state}-(a), where the thick black line is the ground truth of the latent process, the grey lines are the majority vote estimates for each synthetic processes among 
{\color{black}$S=100$} processes,
and the thick blue line and the blue shades are the average latent trajectory of 
{\color{black}$S=100$}
estimates and its one-standard-deviation confidence bands. The estimated trajectories reflect the hidden true state process $\mathcal{Z}(T)$, even at $M=50$. As the number of events increases, our estimation of the latent process becomes more accurate. 

Figure \ref{fig:sim_state}-(b) provides a comparison between the true intensity function for one simulated 50-event process and the estimated intensity function that given the posterior draws of the model parameters and the latent process. The accuracy of the estimated intensity serves as another validation of our estimation of the model parameters and the latent state process. 

For comparison, we include in Figure \ref{fig:sim_state}-(c) the inferred trajectories on the same simulated data sets using MMPP, which suffers from biases and high variability. The inference of the latent process deviates more from the ground truth when events are more bursty in the state 1. This is because MMPP assumes a constant intensity in each state and is not flexible enough to capture the highly heterogeneous event time data that is generated by the MMHP model. As a result, MMPP systematically underestimates the intensity in the state 1 and overestimates the intensity in the state 0. This causes the estimated mean latent process to regress towards 0.5, with the most sizable deviation taking place during the transition period.

\begin{figure}[ht]
 \centering
 \includegraphics[width=0.98\textwidth]{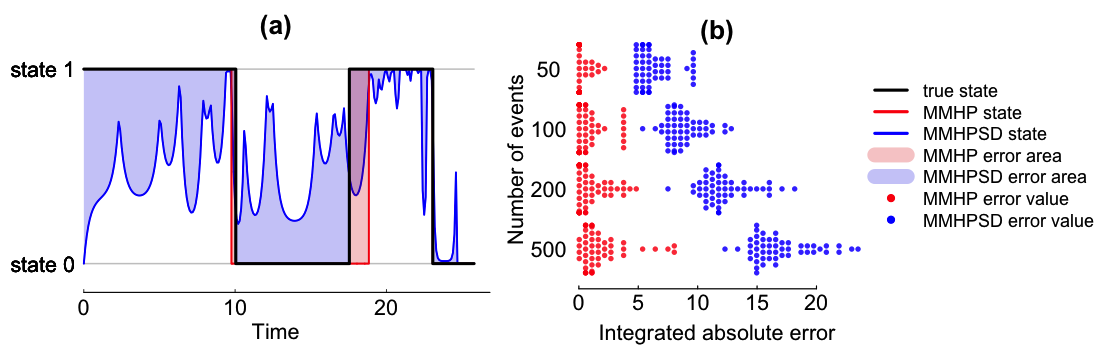}
 \caption{Comparison of integrated absolute error between MMHP and MMHPSD. {\bf (a)} An illustration of the integrated absolute error of a latent process, i.e., $\int_0^T|Z(t)-\hat{Z}(t)|dt$. Here, the black line indicates the true trajectory, the red and blue lines represents the estimated latent process by MMHP and MMHPSD correspondingly, and the red and blue shaded areas correspond to the absolute error of the inferred states using these two methods. {\bf (b)} Comparison of the integrated absolute error between MMHP and MMHPSD under different simulation scenarios.}
 \label{fig:sim_beeswarm}
\end{figure}

We also evaluated the performance of MMHPSD \citep{wang2012markov}, with results shown in Figure \ref{fig:sim_state}-(d). MMHPSD suggests transitions between states that are much more frequent than the ground truth. To quantify this difference between our MMHP model and MMHPSD, we calculate the \textit{integrated absolute error} of the inferred latent process, i.e., $\int_0^T|Z(t)-\hat{Z}(t)|dt$, as shown in Figure \ref{fig:sim_beeswarm}-(a). Figure \ref{fig:sim_beeswarm}-(b) shows the comparison between MMHP and MMHPSD, from which we can see that the overly sensitive state-transition of MMHPSD leads to a larger integrated absolute error.

\subsection{Application to email interactions}
\label{sec:email}

We can now fit our proposed model to the directed email pair
presented in Figure~\ref{fig:ex_email}, consisting of emails from one user to another in a large university over one academic 
semester. In total, several million emails between more than 40k members of the university were collected. Here we consider the single directed pair shown in Figure~\ref{fig:ex_email}, which consists of 78 emails over one academic semester of 122 days.



\begin{figure}[h!]
\centering
\begin{subfigure}{}
   \includegraphics[width=1\textwidth]{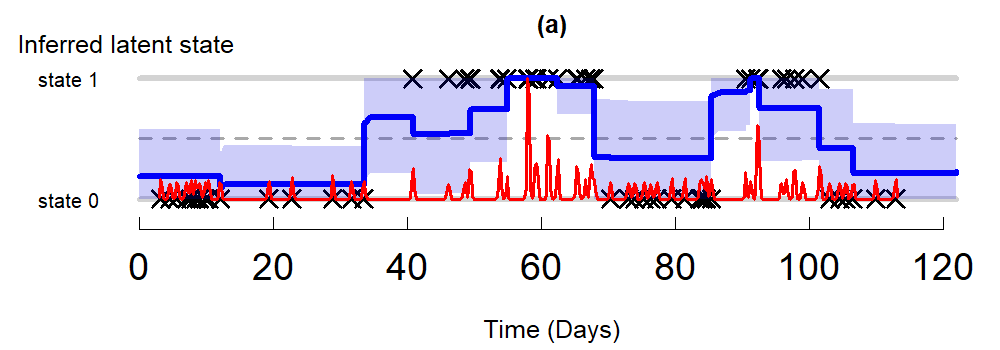}
   \caption*{}
   \label{fig:Ng1} 
\end{subfigure}
\vspace{-1.25cm}
\begin{subfigure}{}
   \includegraphics[width=\textwidth]{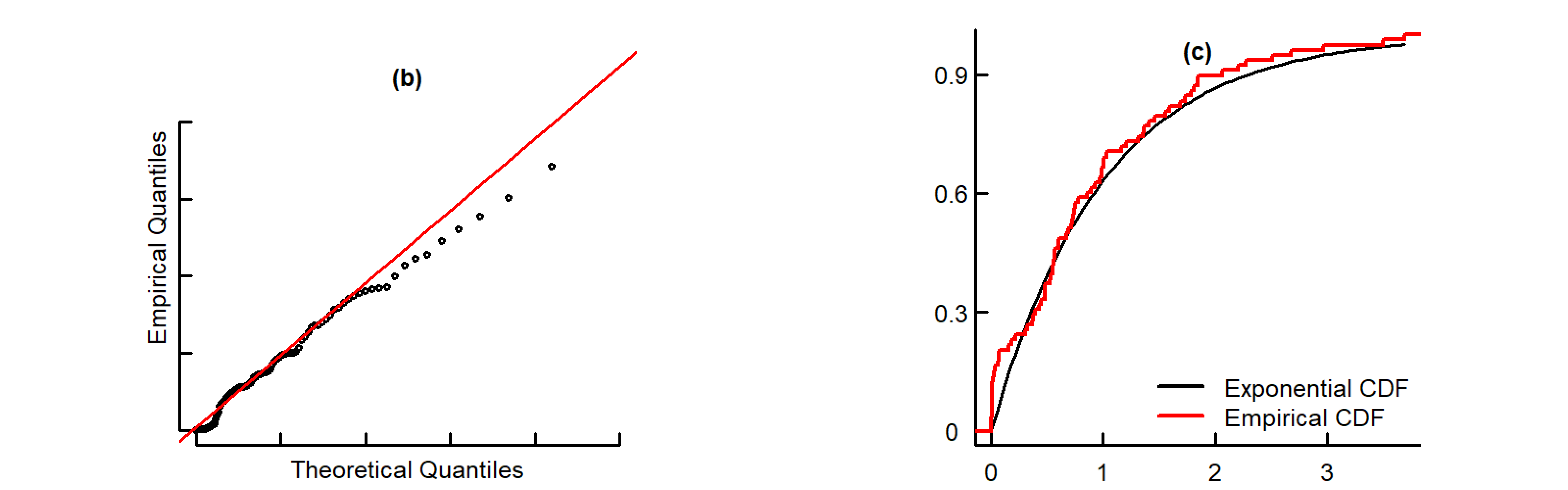}
   \caption*{}
   \label{fig:Ng2}
\end{subfigure}

\caption{Modeling email interaction data. (a) Estimated latent process for one directed pair of emails from Figure~\ref{fig:ex_email}, where the crosses are estimated states at observed event times; blue line is inferred $\mathcal{Z}(T)$ based on the majority vote given posterior draws of latent trajectory; the shaded blue area is the one-standard-error band, constrained to be between $0$ and $1$. KDE estimate overlaid with red line. (b) and (c) show  the QQ plot and KS plot for point process compensators of this directed pair.}
\label{fig:real_one_pair_email}
\end{figure}

We apply the proposed MMHP model to this email interaction data,
using $\mbox{Gamma}(1,1)$ priors for both $\alpha$ and $\beta$, with the other priors being the same as those used in Sec~\ref{sec:mice}.
Figure \ref{fig:real_one_pair_email}-(a) plots the inferred latent states with observed event times and the estimated state trajectory with the one-standard-error band. MMHP separates bursts of many emails over a period of only a few days from less frequent periods, consisting of at most one or two emails per day. This seems well suited to describing behaviour between individuals in this setting, with several periods of frequent contact throughout the semester interspersed between regular but less frequent communication.

A common approach to test the goodness-of-fit of point process models is the time rescaling theorem \citep{brown2002time}. As it states, the \textit{compensators} $\{\Lambda_m:=$ $\int_{t_m}^{t_{m+1}}\lambda(u)du\}_{m=0}^{M-1}$, are expected to be independently distributed following an exponential distribution with rate 1. The MMHP compensators closely follow the exponential distribution with rate 1, as shown in the QQ plot in Figure~\ref{fig:real_one_pair_email}-(b) and Kolmogorov-Smirnov plot in Figure \ref{fig:real_one_pair_email}-(c). Compared to a homogeneous Poisson process, MMPP, MMHPSD and Hawkes process models, the MMHP model gives the best fit in terms of KS statistic. Further comparison with existing models is given in 
Supplement~\ref{supp:checking}.


\subsection{Experiments with social interaction data among mice}
\label{sec:mice}

In \citet{williamson2016temporal}, twelve male mice were placed in a large vivarium at the age of nine weeks. These mice fight each other to establish a social hierarchy. For twenty-one consecutive days, observations were taken during the dark phase of the light cycle when mice are most active. During each observation interval, trained observers recorded the location, time stamp and action of pair-wise mice interactions. For each interaction, the actor (i.e., winner) and recipient (i.e., loser) were recorded. The goal of this study was to understand how mice collectively establish and navigate their social hierarchy over time and to identify inconsistent deviations from a linear order.

Based on our study of the dataset, we conjecture that the social interaction dynamics of mice exhibit two states: {\em active} and {\em inactive}, which can be detected using the proposed MMHP model. After we separate all the interactions into two states, we expect the active state interactions to follow a \textit{linear} hierarchy more closely than with all interactions combined, which suggests an explanation on how social dominance is established among a group of mice. Using multiple measurements of \textit{linearity} for an animal social hierarchy, we show that it is indeed the case. In addition, the inactive state interactions offer insights on social structures among the mice that deviate from the dominance hierarchy. A cluster analysis of the inactive state interactions shows that, as time progresses, the extent of between-cluster interactions decreases, which suggests the social structure may be stabilizing.

From this dataset, we considered the relational event dynamics on a fixed set of actors $V=\{1,2,...,N=12\}$. The data consist of all historic events up to a termination time $T$: $\mathcal{H}(T) = \{(i_m,j_m,t_m)\}_{m=1}^M$. For each directed pair of actors $(i\rightarrow j)$, the sequence of interaction event times is denoted by $\mathcal{H}^{(i\rightarrow j)}(T) = \{t^{(i\rightarrow j)}_m\}_{m=0}^{M^{(i\rightarrow j)}}$, where $t^{(i\rightarrow j)}_0=0$, and $M^{(i\rightarrow j)}$ is the number of interaction events initiated by $i$ and received by $j$. For the purpose of this paper, we treat each observation window separately and independently. This allows us to disregard non-observation time. 

Assume that, for each directed pair $(i\rightarrow j)$, the dyadic events $\mathcal{H}^{(i\rightarrow j)}(T)$ follow the MMHP model. The parameters of each process share the same prior distribution and vary across pairs. To improve the estimation of the state trajectories, we assume that \vspace{-8pt} $$\lambda_0^{(i\rightarrow j)}\sim\mbox{N}^+(\frac{1}{\Delta t^{(i\rightarrow j)}_{\mbox{max}}},0.1),\vspace{-8pt}$$ where $\Delta t^{(i\rightarrow j)}_{\mbox{max}}$ is the maximum over all interevent times for $(i\rightarrow j)$. We assume that the latent state transitions should be less frequent than the fight frequencies, and set $q_0=w_0\lambda_0, q_1=w_1\lambda_1$, where $w_0\sim\mbox{Beta}(0.5,0.5),w_1\sim\mbox{Beta}(0.5,0.5)$. The other parameters share the following priors: 
$$
\alpha\sim \log\mbox{N}(\mu_{\alpha},\sigma_{\alpha}), \beta\sim \log\mbox{N}(\mu_{\beta},\sigma_{\beta}).
$$
These log-Normal priors allow for the longer tails present considering all interactions within a cohort, compared to the single email interaction pair.

\begin{figure}[ht]
    \centering
    \includegraphics[width=\textwidth,height = 5.2cm]{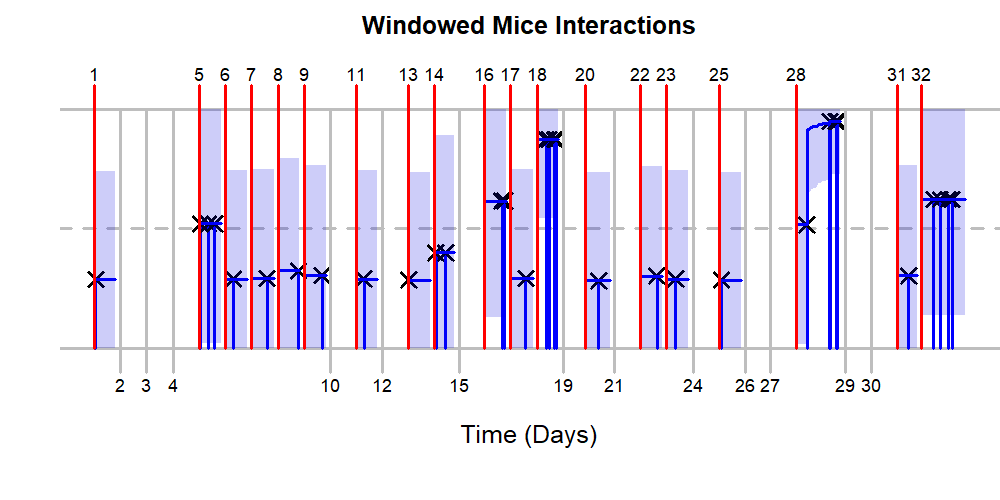}
    \caption{Interactions between two mice over several weeks.
    Here the red lines indicate observation windows (and corresponding days) with interactions while grey lines indicate time periods where no interactions were observed.
    Although we only observe interactions in some of the time windows, we are still able to estimate the latent process in the windows with observed events. This also illustrates the latent state changing within a single time window.}
    \label{fig:mmhp_mice_windows}
\end{figure}

The fights between one pair of these mice are shown in
Figure~\ref{fig:mmhp_mice_windows}. The mice are observed each day, with some observation periods longer than others.
In several of these observation periods no fights occur. 
The continuous time MMHP model is flexible in that it 
allows the latent state to change during longer observation
periods, as seen on day 8 in Figure~\ref{fig:mmhp_mice_windows}. The latent state is not constrained to remained fixed within a given observation window.

\begin{figure}[ht]
 \centering
 \subfigure[All fights]
 {
 \includegraphics[width=0.25\linewidth]{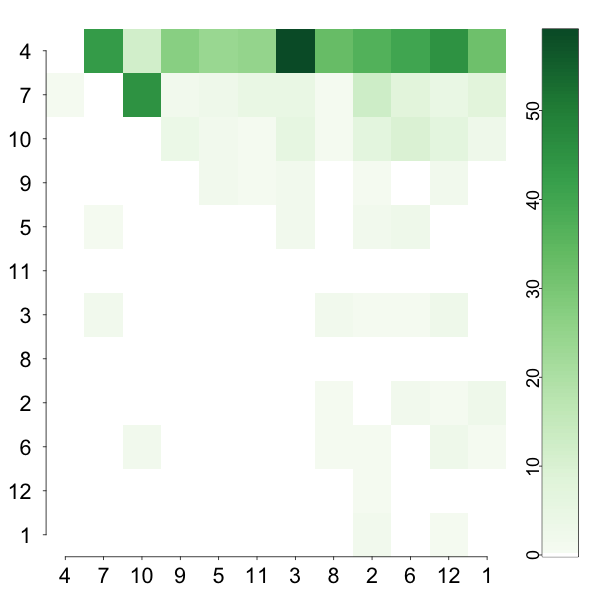}
 \label{fig:state_separation_1}
 }
 \hspace{.3cm}
 \subfigure[Active state]
 {
 \includegraphics[width=0.25\linewidth]{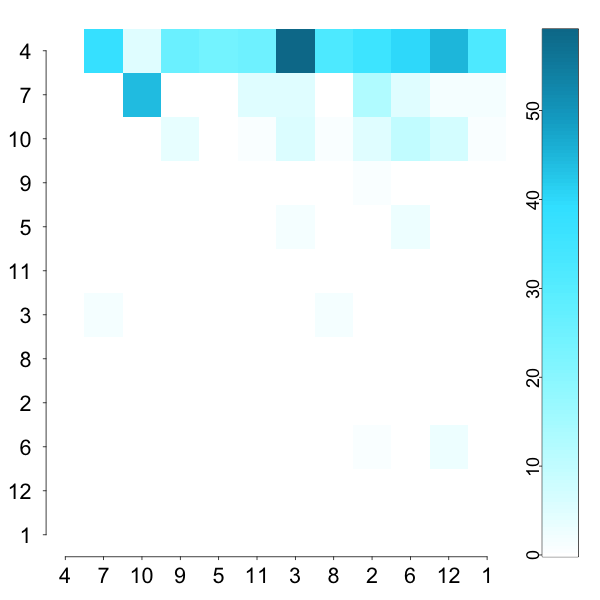}
 \label{fig:state_separation_2}
 }
 \hspace{.3cm}
 \subfigure[Inactive state]
 {
 \includegraphics[width=0.25\linewidth]{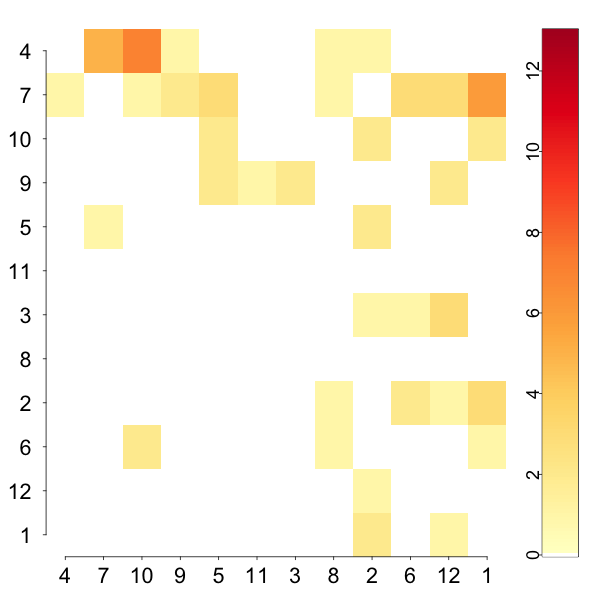}
 \label{fig:state_separation_3}
 }
 \caption{Plots of mice win/loss matrices for all fights and by states (sorted by \textit{I}\&\textit{SI} method \citep{vries1998finding}).}
 \label{fig:state_separation}
\end{figure}

The \textit{active/inactive} state separation by MMHP can also help us understand the inconsistency between mice interaction behaviors and their hierarchy ranks. Given a set of fights among a group of mice, one can calculate the \textit{win/loss} matrix, which is a frequency sociomatrix of wins and losses \citep{so2015social}. The $(i,j)$-th entry in the \textit{win/loss} matrix represents the number of times $i$ won against $j$. For one cohort, Figure \ref{fig:state_separation} plots the win/loss matrices for all fights and by the \textit{active/inactive} state.
{\color{black} There were 929 fights in total in this cohort,
of which 805 were classified as belonging to the active state.}
The order of rows and columns corresponds to the ranks of mice using the \textit{I}\&\textit{SI} method \citep{vries1998finding}. If the interactions strictly follow the social hierarchy, we expect to see all the interactions in the upper triangle with a few exceptions that are close to the diagonal. This is not the case in the overall panel (Figure \ref{fig:state_separation}-(a)). The upper triangular structure in Figure \ref{fig:state_separation}-(b) suggests that our identified active state of bursty and intensive fights agrees with hierarchical rank. The inactive state interactions are deviations from the hierarchy. This suggests that these interactions might be motivated by the mice's need to explore the social hierarchy without intensive engagements. 

\begin{table}[h]
\centering
\begin{tabular}{lp{0.2\linewidth}p{0.2\linewidth}p{0.2\linewidth}}
\hline
Measurement & All fights & Active state & Inactive state \\ \hline
Directional Consistency & 0.96 & 0.99 & 0.85  \\ 
Triangle transitivity & 1.00 & 1.00 & 0.80  \\ 
Inconsistencies in ranking & 3  & 1  & 2 \\ \hline
\end{tabular}
\caption{Measures of social hierarchy linearity in the same cohort}
\label{table:state_separation}
\end{table}

To better quantify how closely a set of interactions follow a linear hierarchy, we calculate the following three measures of social hierarchy linearity for the \textit{win/loss} matrix as in Figure \ref{fig:state_separation}: \textit{Directional consistency} \citep{leiva2008testing}, \textit{Triangle transitivity} \citep{mcdonald2012comparative} and \textit{Inconsistency} in the ranking \citep{vries1998finding}. Given a win/loss matrix $W$, \textit{directional consistency} is defined as $\sum_{i<j}\frac{\max{(W_{ij},W_{ij})}-\min{(W_{ij},W_{ij})}}{W_{ij}+W_{ij}}$, which is the difference in the proportions of fights won by the more dominant individuals and that by the more subordinate individuals. For three individuals $(i,j,k)$, \textit{triangle transitivity} measures the proportion of triad relations satisfying transitivity, i.e. $\mathbbm{1}_{\{i>j,j>k,i>k\}}$, where $i>j$ represents $i$ dominates $j$ and $\mathbbm{1}_{\{\cdot\}}$ is the indicator function. For $\tilde{W}$, which is $W$ with its rows and columns reordered according to a ranking, \textit{inconsistency} in the ranking equals to $\sum_{i>j}\mathbbm{1}_{\{\tilde{W}_{ij}>\tilde{W}_{ji}\}}$. A perfect linear hierarchy ranking would give zero inconsistency. Table \ref{table:state_separation} shows the results of the above measures corresponding to the win/loss matrix shown in Figure \ref{fig:state_separation}. 

\begin{figure}[ht]
 \centering
 \subfigure[State separation measurements]
 {
 \includegraphics[width=0.51\linewidth]{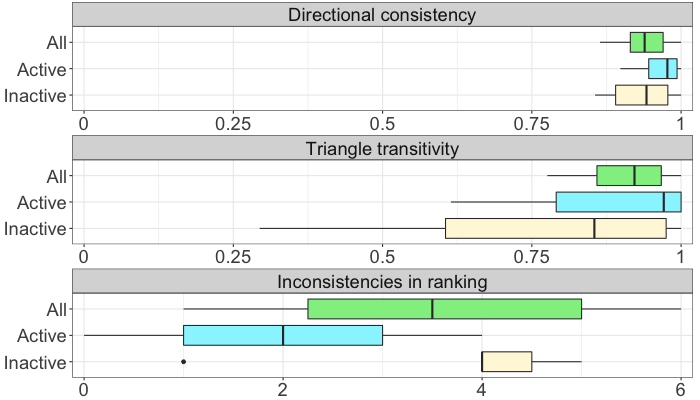}
 }
 \subfigure[Inactive state cluster trend]
 {
 \includegraphics[width=0.44\linewidth]{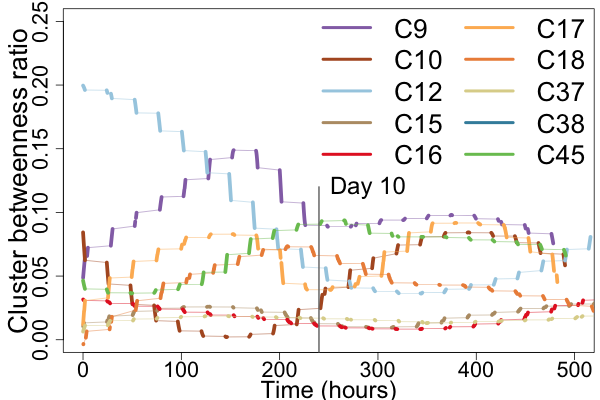}
 }
 \caption{Summary of state separation result in 10 cohorts.}
 \label{fig:real_cohorts}
\end{figure}

Figure \ref{fig:real_cohorts}-(a) shows the boxplots of these three measurements for ten cohorts, which are calculated using all, active, and inactive interactions. All three measurements suggest a stronger linear hierarchy among the active-state interactions comparing to 
all interactions combined.

\citet{williamson2016temporal} found that day 5 and day 10 were the start and end of social hierarchy stabilization. 
Figure \ref{fig:real_cohorts}-(b) summarized this trend for all cohorts: each line in the plot indicates a cohort's evolving ratio of between-cluster interactions and within-cluster interactions. The thicker line segments represent the ratio values calculated for each day by using interactions within a sliding one-hour time window. Due to the fact that the study has different observation times within a day across cohorts, we align all the cohorts by interpolating the calculated segments (thick lines) with a thin line. We see that all cohorts show stable clustering after the 10th day.

\section{Discussion and conclusion}
\label{Conclusion}

Although dynamic social interactions have been widely studied recently, existing models cannot adequately capture the patterns of event dynamics in social interaction data -- sporadic with bursts and long wait time. 
In this paper, we proposed a Markov Modulated Hawkes Process (MMHP) model and its inference algorithms, which segments different dynamic patterns in event arrival times data. Results from numerical experiments provide validating evidence on the advantages of the proposed method over comparable existing methods, in both simulated studies and a real application to both animal behavior data and email data.

In this paper, the MMHP model and its inference framework are built under the 
assumption that the latent Markov process has two states for better model 
interpretability. It {\color{black}can readily}
be extended to a $R$-state latent Markov process by
modifying the computation of the forward-backward algorithm, {\color{black} 
as we have included in Appendix~\ref{k-state-algo}.
However,
it is challenging to make this framework computationally
practical for $R>2$. This also raises
identifiability concerns between the underlying
latent process and the Hawkes processes. \citet{veen2008estimation} pointed out that the log likelihood function of Hawkes processes may be multi-modal and locally flat. This becomes more notable when we consider $R$ generally parametrized Hawkes processes, modularized by a Markov process.
One potential avenue is to utilise other inference
schemes, such as Markov Jump processes \cite{rao_teh_13}. Initial
simulation studies (not shown here) indicate this is a promising
direction, however this procedure struggles to infer a stable
latent process in our model for $R=2$.
Further work is therefore required to investigate this
inference procedure in this setting before considering the 
general $R$-state problem.
}

MMHP uses the classic exponential kernel in Hawkes processes, where $\alpha$ represents the influence of historical events on the intensity and $\beta$ represents the decay rate of such influence. A potential extension of the proposed model could be introducing covariates into $\alpha$ and $\beta$. Subsequent inference can be easily conducted under our framework. Although our current work emphasizes more on the model interpretation due to the motivation from the social interaction data, our model can also make predictions on the expected number of future events with measures of uncertainty.

{\color{black}
In this paper, we focused on modeling the sporadic dynamic of 
one sequence of event history that can be assumed as one 
non-stationary point process. In practice, some event streams 
might be driven by multiple dynamic processes. One such example 
is \cite{du2015dirichlet},
who fitted a nonparametric Dirichlet mixture of 
Hawkes processes to a large collection of document streams and 
clustered these news articles into separate streams, with each 
stream represented by a Hawkes process. A similar mixture of 
Hawkes processes is considered by \cite{xu2017dirichlet}.
This mixture of 
Hawkes process is able to capture the self excitatory patterns 
seen in mixtures of document streams, but is not able to capture
the long wait times observed in social interaction data that we 
consider here. Adding an inactive state with sporadic event arrivals to such mixture models via Markov modulation is a future direction worth exploring.
}

Furthermore, we assumed independence structure when applying the 
model to a network of animals. However, more research can be carried
out in terms of introducing a dependence structure among different 
animal pairs in a network. Such a network-structured MMHP will lead 
to improvements in model estimates and interpretability. On the 
other hand, combined with network models, MMHP provides a way of 
inferring network structure based on continuous-time event data.

\newpage

\bibliographystyle{apalike}
\bibliography{mmhp_ref}

\newpage

\appendix

\section{Algorithms}
\label{appendix:algo}

\subsection{Posterior Samples}
\label{appendix:post}
Here we provide the algorithm use to draw samples from the posterior of the proposed MMHP model. 

\begin{algorithm}[ht]
 \caption{Posterior sampling of parameters for MMHP.}
 \label{algo:1}
 \begin{algorithmic}
 \Inputs $M$(number of events), $\Delta t_{1:M}$(interevent time)
 \Prior $\delta_0\sim U(0,1),\alpha \sim \mbox{N}(0,5), \beta\sim\log\mbox{N}(0,0.5),$\\ 
 \hspace{14pt}$\lambda_1\sim\log\mbox{N}(0,1), \lambda_0<\lambda_1, q_0, q_1 \sim\log\mbox{N}(-1,1).$
 \Likelihood \\
 \hspace{10pt}\textbf{Initialize}:
 $m\gets 0$; 
 $\mathcal{A}_m[1]\gets(1-\delta_{0})$; 
 $\mathcal{A}_m[2]\gets\delta_{0}$.
 \While{$m<M$}
 \State $m \gets m+1$
 \State $\mathcal{A}_m[1]\gets
 \mathcal{A}_{m-1}[1]\times 
 \tilde{P}_\Theta(Z_m=1, \Delta t_m|Z_{m-1}=1, \Delta t_{1:m-1})+$\\
 $\hspace{66pt}
 \mathcal{A}_{m-1}[2]\times 
 \tilde{P}_\Theta(Z_m=1, \Delta t_m|Z_{m-1}=0, \Delta t_{1:m-1})$
 \State $\mathcal{A}_m[2]\gets
 \mathcal{A}_{m-1}[1]\times 
 \tilde{P}_\Theta(Z_m=0, \Delta t_m|Z_{m-1}=1, \Delta t_{1:m-1})+$\\
 $\hspace{66pt}
 \mathcal{A}_{m-1}[2]\times 
 \tilde{P}_\Theta(Z_m=0, \Delta t_m|Z_{m-1}=0, \Delta t_{1:m-1})$
 \EndWhile
 $P_{\Theta}(\Delta t_{1:M})\gets \mathcal{A}_M[1]+\mathcal{A}_M[2]$
 \EndLikelihood
 \Posterior Use MCMC to sample from posterior distribution.
 \Outputs Posterior draws of $\Theta$.
 \end{algorithmic}
  \end{algorithm}

\subsection{The Viterbi algorithm}
\label{appendix:vit}
Algorithm~\ref{algo:2} describes the Viterbi algorithm discussed in the main text to infer the latent states of MMHP.

\begin{algorithm}[ht]
 \caption{Viterbi algorithm for Markov Modulated Hawkes Process}
 \label{algo:2}
 \begin{algorithmic}
 \Inputs $\hat{\Theta}$(Estimation of parameters), $M$(number of events), $\Delta t_{1:M}$(interevent time)
 \Initialize
 $m\gets 0$; 
 $v_m[1]\gets\log{(1-\hat{\delta}_0)}$; 
 $v_m[0]\gets\log{(\hat{\delta}_0)}$
 \While{$m<M$}
 \State \hspace{1.2em} $m \gets m+1$
 \vspace{-0.5em}
 \begin{flalign*}
 &\hspace{-1.7em}v_m[1]\gets \max_{k}v_{m-1}[k]+
 \log{\tilde{P}_{\hat{\Theta}}(Z_m=1, \Delta t_m|Z_{m-1}=k,\Delta t_{1:m-1})}\\
 &\hspace{-1.7em}v_m[0]\gets \max_{k}v_{m-1}[k]+
 \log{\tilde{P}_{\hat{\Theta}}(Z_m=0, \Delta t_m|Z_{m-1}=k,\Delta t_{1:m-1})}\\
 &\hspace{-1.7em}b_m[1]\gets \argmax\limits_{k}v_{m-1}[k]+
 \log{\tilde{P}_{\hat{\Theta}}(Z_m=1, \Delta t_m|Z_{m-1}=k,\Delta t_{1:m-1})}\\
 &\hspace{-1.7em}b_m[0]\gets \argmax\limits_{k}v_{m-1}[k]+
 \log{\tilde{P}_{\hat{\Theta}}(Z_m=0, \Delta t_m|Z_{m-1}=k,\Delta t_{1:m-1})} 
 \end{flalign*}
 \vspace{-1.5em}
 \EndWhile
 \State $z_M^*\gets\argmax\limits_kv_M[k]$
 \While{$m^\prime\leq M$}
 \State $z_{M-m^\prime}^*\gets b_{M-m^\prime+1}[z_{M-m^\prime+1}^*]$
 \EndWhile
 \Outputs Global optimal sequence of latent state $(z_0^*,z_1^*,...,z_M^*)$
 \end{algorithmic}
 \end{algorithm}

\subsection{The forward backward algorithm}

\label{appendix:for_back}
The forward-backward algorithm is a dynamic programming algorithm for computing the marginal likelihood of a sequence of observations from complete-data likelihood, by iteratively marginalizing out hidden state variables. As shown in Section \ref{inference-variation}, we obtained the approximated likelihood after marginalizing over $\mathcal{Z}(T)\in\mathcal{S}_{\tilde{\mathcal{Z}}(T)}$, i.e. $\tilde{P}_\Theta(\mathcal{H}(T),\tilde{\mathcal{Z}}(T))$. This is a function of $\tilde{\mathcal{Z}}(T)=\{Z_0, Z_1, ...., Z_M\}$, which are the states at event times. Utilizing the forward variable, we can then marginalize over $\tilde{\mathcal{Z}}(T)$ computationally efficiently. For convenience, we denote the event history up to $m$-th event $\mathcal{H}(t_m)$ through their interevent times $\Delta t_{1:m}:=\{\Delta t_i := t_i-t_{i-1}, i=1,...,m\}$. Hence the forward variable can be defined as $\mathcal{A}_m(z_m,\Delta t_{1:m}):= \tilde{P}_{\Theta}(Z_m=z_m,\Delta t_{1:m})$. It satisfies the initial condition that $\mathcal{A}_0(z_0)=P(Z_0=z_0)$. The forward iteration can be derived as
\begin{align*}
 & \mathcal{A}_m(z_m,\Delta t_{1:m}) \\
 &= \tilde{P}_{\Theta}(Z_m=z_m,\Delta t_{1:m})\\
 &= \sum_{z_{m-1}}\tilde{P}_{\Theta}(Z_{m-1}=z_{m-1},\Delta t_{1:m-1})\tilde{P}_{\Theta}(Z_m=z_m, \Delta t_{1:m}|Z_{m-1}=z_{m-1},\Delta t_{1:m-1})\\
 &= \sum_{z_{m-1}}\mathcal{A}_{m-1}(z_{m-1},\Delta t_{1:m-1})\tilde{P}_{\Theta}(Z_m=z_m, \Delta t_m|Z_{m-1}=z_{m-1},\Delta t_{1:m-1})
\end{align*}

Given the last forward variable, we obtain the whole marginalized likelihood as $$\sum_{\mathcal{Z}(T)}\tilde{P}_\Theta(\mathcal{H}(T),\tilde{\mathcal{Z}}(T))=\sum_{z_M}\mathcal{A}_M(z_M,\Delta t_{1:M})$$

{\color{black}

\section{General R State Algorithm}
\label{k-state-algo}
For completeness, we describe our MMHP
procedure for general $R$ states here. For this, we 
define CTMC with R states as $Z(t), t\in[0,T], Z(t)\in\{1,2,...,R\}$. The initial 
probability vector is $\mathbb{\delta}:=\{\delta_1, ..., \delta_R\}$, where 
$\sum_{r=1}^R\delta_r=1$. The infinitesimal generator matrix $Q$ is a $R\times K$ 
matrix defined as

\begin{align*}
Q &  = 
\begin{bmatrix}
q_{1,1} & q_{1,2} & \cdots & q_{1,R} \\
q_{2,1} & q_{2,2} & \cdots & q_{2,R} \\
\vdots  & \vdots  & \ddots & \vdots  \\
q_{R,1} & q_{R,2} & \cdots & q_{R,R}
\end{bmatrix},
\end{align*}

where $q_{r,r}=-\sum_{j\neq r} q_{r,j}$.

For each $t\geq 0$, there is a probability transition matrix, denoted as 
$\mathbf{P}(t):=$ $[\mathbf{P}_{ij}(t)]_{i,j\in\{0,1\}}$. Each entry 
$\mathbf{P}_{ij}(t)$ is defined as the probability that the chain will be in state $j$
at time $u+t$ ($t>0$) given the chain is in state $i$ at time $u$, i.e. for each 
$u\geq 0$, $$\mathbf{P}_{ij}(t)=P(Z(u+t)=j|Z(u)=i).$$ Following Kolmogorov's forward 
equation, we know that $\mathbf{P}(t) = \exp{(Qt)}$.

Given $Z(t)=r$, define the Hawkes intensity as 
$\lambda_{r}(t)=\lambda_{0,r}+\alpha_r\sum_{t_m<t}e^{-\beta_r(t-t_m)}$.

To derive the forward algorithm:
\begin{align*}
 & \mathcal{A}_m(z_m,\Delta t_{1:m}) \\
 &= \tilde{P}_{\Theta}(Z_m=z_m,\Delta t_{1:m})\\
 &= \sum_{z_{m-1}}\tilde{P}_{\Theta}(Z_{m-1}=z_{m-1},\Delta t_{1:m-1})\tilde{P}_{\Theta}(Z_m=z_m, \Delta t_{1:m}|Z_{m-1}=z_{m-1},\Delta t_{1:m-1})\\
 &= \sum_{z_{m-1}}\mathcal{A}_{m-1}(z_{m-1},\Delta t_{1:m-1})\tilde{P}_{\Theta}(Z_m=z_m, \Delta t_m|Z_{m-1}=z_{m-1},\Delta t_{1:m-1})
\end{align*}

Here, to be precise,
\begin{align*}
& \tilde{P}_{\Theta}(Z_m=z_m, \Delta t_m|Z_{m-1}=z_{m-1},\Delta t_{1:m-1}) \\
&= \mathbf{P}_{z_{m-1},z_{m}}(\Delta t_m) \lambda_{z_m}(t_m|\mathcal{H}(t_m))\exp{\{-\int_{t_{m-1}}^{t_m}\lambda(u|\mathcal{H}(u)) du\}}.
\end{align*}

We give the full details of the algorithms required for the
general R-state model in Algorithm~\ref{algo_k:1} and 
Algorithm~\ref{algo_k:2}.

\begin{algorithm}[h!]
 \caption{Posterior sampling of parameters for R-state MMHP.}
 \label{algo_k:1}
 \begin{algorithmic}
 \Inputs $M$(number of events), $\Delta t_{1:M}$(interevent time)
 \Prior $\delta_0\sim U(0,1),\alpha \sim \mbox{N}(0,5), \beta\sim\log\mbox{N}(0,0.5),$\\ 
 \hspace{14pt}$\lambda_r\sim\log\mbox{N}(0,1), Q \sim\log\mbox{N}(-1,1).$
 \Likelihood \\
 \hspace{10pt}\textbf{Initialize}:
 $m\gets 0$; 
 $\mathcal{A}_m[r]\gets\delta_r$, for $r\in\{1,...,R\}$.
 \While{$m<M$}
 \State $m \gets m+1$
 \For{r=1,...,R}
 \State $\mathcal{A}_m[r]\gets
 \sum_{j=1}^R \mathcal{A}_{m-1}[j]\times 
 \tilde{P}_\Theta(Z_m=r, \Delta t_m|Z_{m-1}=j, \Delta t_{1:m-1})$
 \EndFor
 \EndWhile
 $P_{\Theta}(\Delta t_{1:M})\gets \sum_{r=1}^R\mathcal{A}_M[r]$
 \EndLikelihood
 \Posterior Use MCMC to sample from posterior distribution.
 \Outputs Posterior draws of $\Theta$.
 \end{algorithmic}
\end{algorithm}

\begin{algorithm}[h!]
 \caption{Viterbi algorithm for Markov Modulated Hawkes Process}
 \label{algo_k:2}
 \begin{algorithmic}
 \Inputs $\hat{\Theta}$(Estimation of parameters), $M$(number of events), $\Delta t_{1:M}$(interevent time)
 \Initialize
 $m\gets 0$; 
 $v_m[r]\gets\log{(\hat{\delta}_r)}, r=1,...,R$.
 \While{$m<M$}
 \State \hspace{1.2em} $m \gets m+1$
 \For{r=1,...,R}
 \State $v_m[r]\gets \max_{j}v_{m-1}[j]+
 \log{\tilde{P}_{\hat{\Theta}}(Z_m=r, \Delta t_m|Z_{m-1}=j,\Delta t_{1:m-1})}$
 \State $b_m[r]\gets \argmax\limits_{j}v_{m-1}[j]+
 \log{\tilde{P}_{\hat{\Theta}}(Z_m=r, \Delta t_m|Z_{m-1}=j,\Delta t_{1:m-1})}$
 \EndFor
 \EndWhile
 \State $z_M^*\gets\argmax\limits_r v_M[k]$
 \While{$m^\prime\leq M$}
 \State $z_{M-m^\prime}^*\gets b_{M-m^\prime+1}[z_{M-m^\prime+1}^*]$
 \EndWhile
 \Outputs Global optimal sequence of latent state $(z_0^*,z_1^*,...,z_M^*)$
 \end{algorithmic}
\end{algorithm}

}


\newpage


\newcommand{\beginsupplement}{%
        \setcounter{table}{0}
        \renewcommand{\thetable}{S\arabic{table}}%
        \setcounter{figure}{0}
        \renewcommand{\thesection}{S\arabic{section}}
        \setcounter{section}{0}
        \renewcommand{\thefigure}{S\arabic{figure}}%
     }
\beginsupplement

\section*{Supplementary Material}

\renewcommand\thefigure{S\arabic{figure}}
\setcounter{figure}{0}  

\renewcommand{\thesection}{S\arabic{section}}
\setcounter{subsection}{0}

\subsection{Discussion of data binning}
\label{supp:data_binning}

\begin{figure}[ht]
 \centering
 \includegraphics[width=\textwidth]{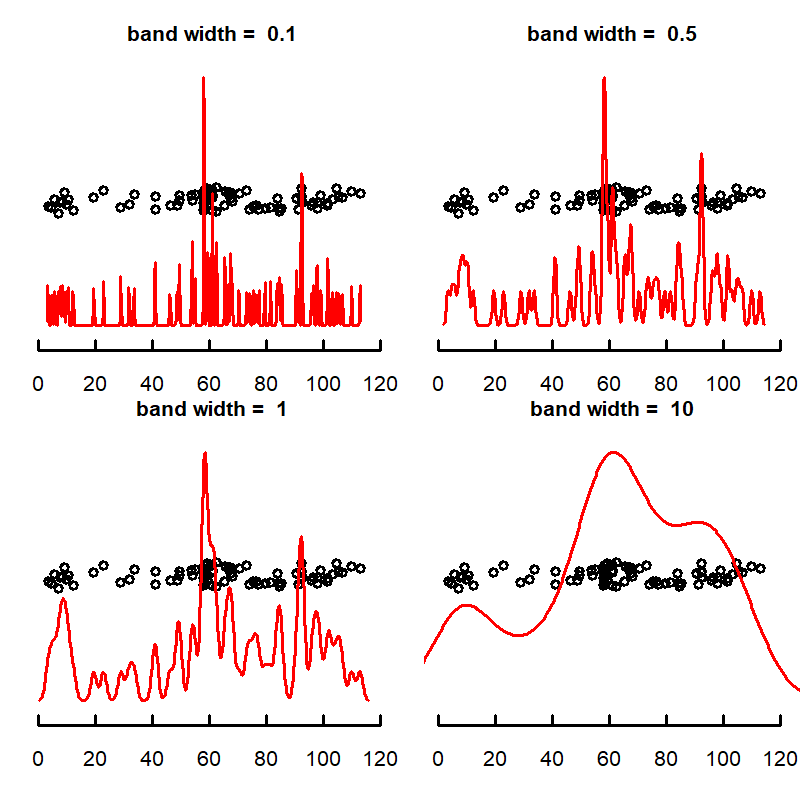}
 \caption{Plots of the jittered event times for the email interaction data along with corresponding kernel density estimates for a range of decreasing bandwidths.}
 \label{fig:smoothed_grid}
\end{figure}

To further illustrate the difficulties of correctly binning continuous time data, we illustrate kernel density estimates with a range of bandwidths. As can be seen in Figure~\ref{fig:smoothed_grid}, it is only when the bandwidth of the kernel density is sufficiently small that it begins to capture the sporadic and bursty nature of interaction times. This indicates that a continuous time model is needed to fully capture the dynamics in this data.

\subsection{Further model comparison}
\label{supp:checking}
Here we illustrate further how existing models fail to capture the latent dynamics present in the data we analysed. \par 

\begin{figure}[h!]
 \centering
 \includegraphics[width=\textwidth]{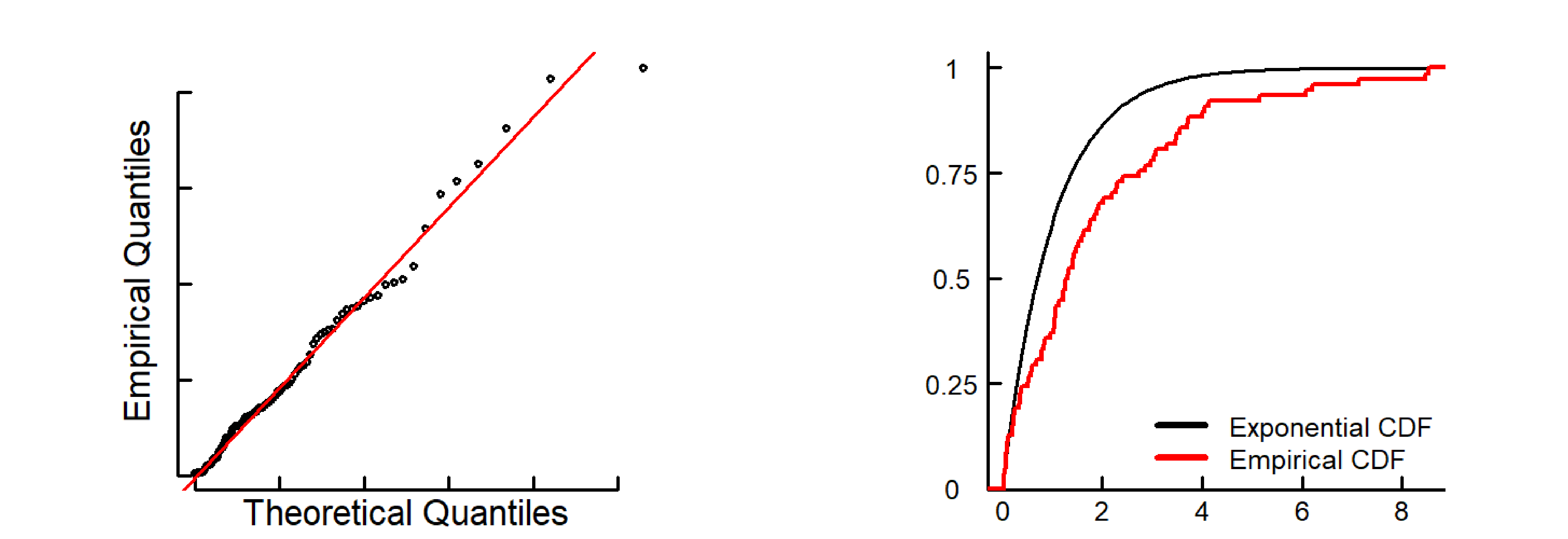}
 \caption{Compensator plots for Hawkes. These illustrate how this model is not able to capture the latent dynamics present in the data.}
 \label{fig:hawkes_comp}
\end{figure}

\begin{figure}[h!]
 \centering
 \includegraphics[width=\textwidth]{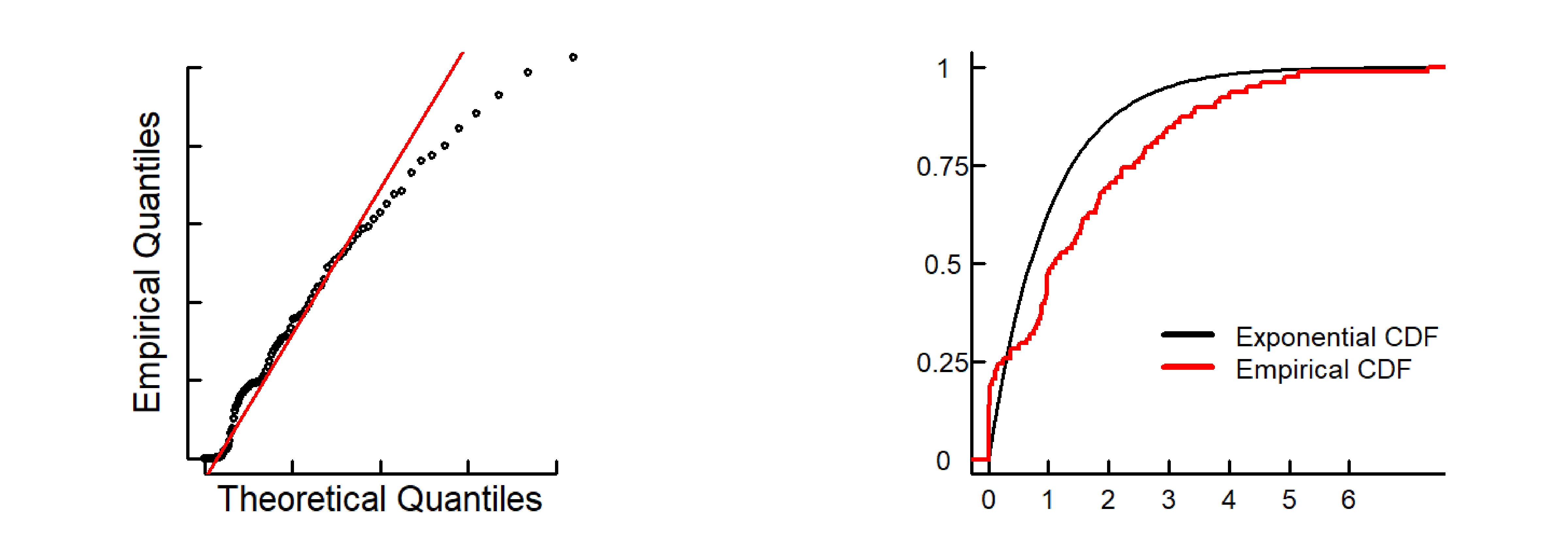}
 \caption{Compensator plots for MMPP. These illustrate how this model is not able to capture the latent dynamics present in the data.}
 \label{fig:mmpp_comp}
\end{figure}

\begin{figure}[h!]
 \centering
 \includegraphics[width=\textwidth]{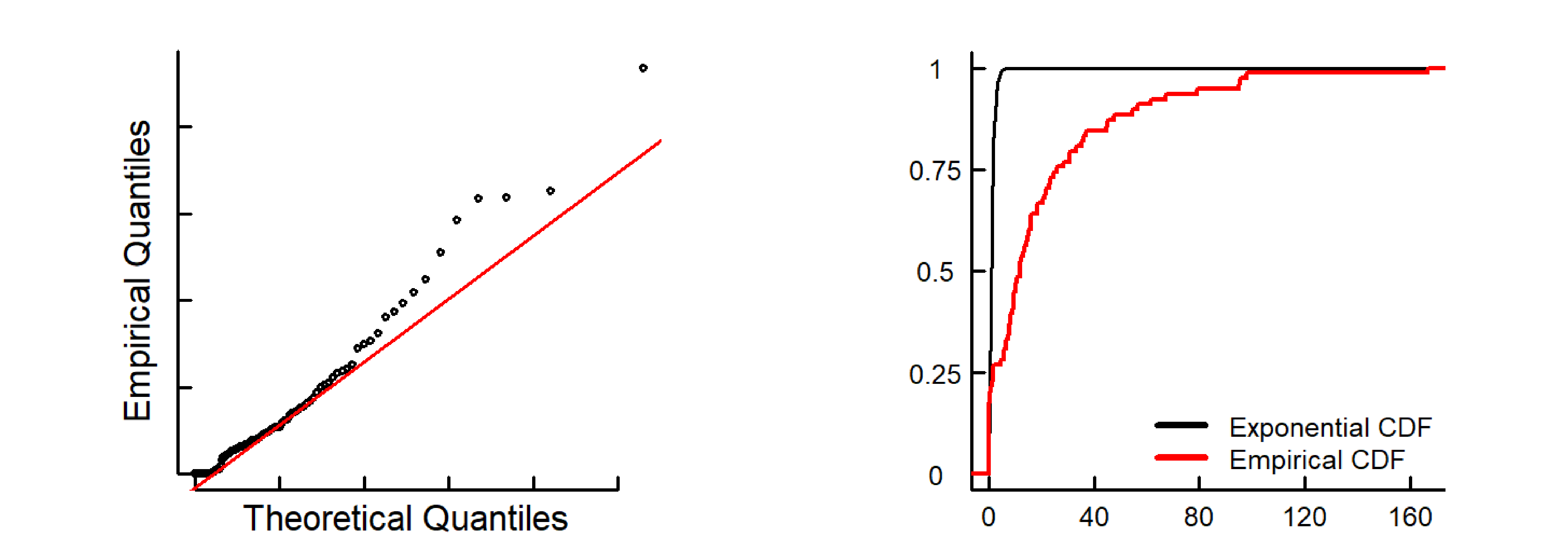}
 \caption{Compensator plots for MMHPSD. These illustrate how this model is not able to capture the latent dynamics present in the data.}
 \label{fig:mmhpsd_comp}
\end{figure}

We fit each of the other models from the literature to the email interaction data and illustrate lack of model fit using the methods described in Sec~\ref{sec:email}. Figures~\ref{fig:hawkes_comp},\ref{fig:mmpp_comp} and \ref{fig:mmhpsd_comp} so model evaluation tools for each of the alternative models considered in the paper. As shown by the compensator plots, each of these models are not able to capture the nature of this arrival data. Similarly, performing a Kolmogorov-Smirnov test on the rescaled inter-event times under each model indicates departures from independent identically distributed exponential random variables.


\end{document}